\documentclass[aps,prc,superscriptaddress,showpacs,amsmath,amssym,
nofootinbib,
twocolumn]{revtex4-1}

\usepackage{array}
\usepackage{graphicx}
\usepackage{color}
\usepackage{bm}
\usepackage[sort&compress]{natbib}

\newcommand{\beq}{\begin{equation}}
\newcommand{\eeq}{\end{equation}}
\newcommand{\etal}{\emph{et al.}}
\begin{document}

\title{Strong plasma screening in thermonuclear reactions: Electron drop model}
\author{P.\ A.\ Kravchuk}
\affiliation{St.~Petersburg State Polytechnical University,
Politekhnicheskaya 29, St.~Petersburg 195251, Russia}
\author{D.\ G.\ Yakovlev}
\affiliation{Ioffe Physical Technical Institute, Politekhnicheskaya
26, St.~Petersburg 194021, Russia}

\date{\today}

\begin{abstract}
We analyze enhancement of thermonuclear fusion reactions due to
strong plasma screening in dense matter using a simple electron drop
model. The model assumes fusion in a potential that is screened by an
effective electron cloud around colliding nuclei (extended Salpeter
ion-sphere model). We calculate the mean field screened Coulomb
potentials for atomic nuclei with equal and nonequal charges,
appropriate astrophysical $S$ factors, and enhancement factors of
reaction rates. As a byproduct, we study analytic behavior of the
screening potential at small separations between the reactants. In
this model, astrophysical $S$ factors depend not only on nuclear
physics but on plasma screening as well. The enhancement factors are
in good agreement with calculations by other methods. This allows us
to formulate the combined, pure analytic model of strong plasma
screening in thermonuclear reactions. The results can be useful for
simulating nuclear burning in white dwarfs and neutron stars.
\end{abstract}

\maketitle

\section{Introduction}
\label{s:introduct}

It is well known that nuclear reactions in compact stars, which
contain matter of high density, can be strongly modified by plasma
physics effects (e.g., Ref.\ \cite{svh69}). Under compact stars we
mean white dwarfs and neutron stars \cite{st83}. Central densities of
massive white dwarfs can be as high as $10^{10}$ g~cm$^{-3}$.
Carbon/oxygen burning in the cores of white dwarfs is thought to
trigger type Ia supernova explosions (e.g., Refs.\
\cite{hoeflich06,vankerkwijketal2010}). Neutron stars contain the
outer and inner crust, where atomic nuclei are available and can
participate in various reactions which result in steady-state and
explosive burning and nucleosynthesis
\cite{schatz03,cummingetal05,Brown06}. For instance, we can mention
helium or carbon burning leading to type I X-ray bursts or
superbursts; these reactions occur at densities $\lesssim 10^{10}$
g~cm$^{-3}$.

The plasma physics effects modify the reactions at sufficiently high
densities and not very high temperatures when the plasma of atomic
nuclei becomes strongly non-ideal due to strong Coulomb coupling. At
these conditions, the well known classical thermonuclear burning
regime \cite{bbfh57} is no longer valid. With increasing density
and/or decreasing temperature, one has a sequence of four other
nuclear burning regimes \cite{svh69} which are: thermonuclear burning
with strong plasma screening; intermediate thermo-pycnonuclear
burning; thermally enhanced pycnonuclear burning; and
temperature-independent pycnonuclear burning.

In this paper we address the thermonuclear burning with strong plasma
screening, which is realized in a wide range of temperatures and
densities of matter and is important for applications in compact
stars. In this regime, the plasma of atomic nuclei (ions) is strongly
coupled but mostly classical (quantum effects in motion of ions are
weak).
The plasma effects are well known to enhance thermonuclear
reaction rates and are conveniently described by the enhancement
factor $f$ ($f \geq 1$)
\begin{equation}
  f=R/R_0,
\label{e:enhanc}
\end{equation}
where $R$ is the actual rate and $R_0$ is the rate calculated
neglecting the plasma screening. Unless the contrary is indicated,
subscript 0 will mark quantities calculated neglecting the screening.
The factor $f$ will be the basic quantity of our interest. It has
been calculated using different techniques and approximations in a
number of publications cited in Sec.\ \ref{s:poten}, starting from
the seminal paper by Salpeter \cite{Salp}.

Our aim here is to consider a simple model for strong plasma
screening in thermonuclear reactions. In Sec.\ \ref{s:param} we
outline physical conditions in dense stellar matter. In Sec.\
\ref{s:model} we formulate the model. Then we consider the mean
plasma screening potentials (Sec.\ \ref{s:poten}), the basic
Salpeter's model for plasma screening in thermonuclear reactions
(Sec.\ \ref{s:Salpeter}), as well as astrophysical $S$ factors (Sec.\
\ref{s:sfact}) and enhancement factors (Sec.\ \ref{s:rate}) for our
model. In Sec.\ \ref{s:discussion} we discuss our main results and
propose the combined analytic model for strong plasma screening in
thermonuclear regime; we conclude in Sec.\ \ref{s:conclusions}. Some
technical details are presented in the Appendices.

\section{Plasma parameters}
\label{s:param}

Atomic nuclei in dense stellar matter  are fully ionized by huge
electron pressure, and the electrons are so energetic that constitute
almost rigid background of negative charge in which the ions move.
Generally, we have multi-component ion mixture because we study
nuclear fusion reactions involving equal or different nuclei, and the
reaction products (daughter nuclei) are also present there. We
consider a mixture of ion species $j=1,2,\ldots$, with atomic numbers
$A_j$ and charge numbers $Z_j$. Let $n_j$ be the number density of
ions $j$. The total number density of ions is $n=\sum_j n_j$; the
electron number density is $n_e = \sum_j Z_j n_j$.

It is convenient to introduce the Coulomb coupling parameter
$\Gamma_j$ for ions $j$
(e.g., Ref.\ \cite{hpy07}),
\begin{eqnarray}
 &&  \Gamma_j={ Z_j^2 e^2 \over a_j k_{B} T}
   ={Z_j^{5/3}e^2 \over a_e k_{B} T} ,
\label{Gammaj} \\
 &&  a_e=\left( 3 \over 4 \pi n_e \right)^{1/3},
   \quad
   a_j=Z_j^{1/3}a_e,
\nonumber
\end{eqnarray}
where $T$ is the temperature, $k_{B}$ is the Boltzmann constant,
$a_e$ is the electron-sphere radius, and $a_j$ is the ion-sphere
radius (for a sphere around a given ion, where the electron charge
compensates the ion charge). Therefore, $\Gamma_j$ is the ratio of a
typical electrostatic energy of the ion to the thermal energy. If
$\Gamma_j \ll 1$ then the ions constitute an almost ideal Boltzmann
gas, while for $\Gamma_j \gtrsim 1$ they are strongly coupled by
Coulomb forces (constitute either Coulomb liquid or solid).

The strongly coupled plasma is accurately described as an ensemble of
closely packed ion spheres. The Coulomb energy of the ion sphere
(including the electrostatic energy of the electron cloud and the
energy of electron-ion interaction) is
\begin{equation}
   W(Z)=-\frac{9}{10}\,\frac{e^2 Z^2}{a_j}=
   -\frac{9}{10}\,\frac{e^2Z^{5/3}}{a_e}.
\label{e:W(Z)}
\end{equation}

Let us consider a fusion reaction $(A_1,Z_1)+(A_2,Z_2)\to
(A_{c},Z_{c})$, with $A_{c}=A_1+A_2$ and $Z_{c}=Z_1+Z_2$ (subscript
$c$ refers to a compound nucleus), and introduce the parameters
\begin{equation}
  a_c=a_eZ_c^{1/3},\quad a_{12}=\frac{a_1+a_2}{2},
  \quad E_{12}=\frac{Z_1 Z_2 e^2}{a_{12}},
\label{a12}
\end{equation}
\begin{equation}
 \Gamma_{12}=\frac{E_{12}}{k_BT}, \quad
 \tau= \left( 27 \pi^2 \mu \, Z_1^2\, Z_2^2 e^4 \over
  2 k_{B} T \hbar^2 \right)^{1/3},\quad
  \zeta=\frac{3\Gamma_{12}}{\tau}.
\label{tau12}
\end{equation}
Here, $\mu=m_1 m_2/(m_1+m_2)$ is the reduced mass of the reactants,
$E_{12}$ is a convenient unit of their electrostatic energy, $\tau$
is the basic parameter of thermonuclear reactions, $\zeta$ measures
the importance of quantum effects in ion motion for thermonuclear
reactions. In stellar matter one usually has $\tau \gg 1$;
$\exp(-\tau)$ determines the probability of quantum tunneling through
the Coulomb barrier neglecting plasma screening. Let $E$ be a center
of mass energy of the nuclei. Then the typical (Gamow-peak) energy
which contributes to the thermonuclear reaction neglecting plasma
screening is $E_{p0}= k_BT\tau/3$. Introducing convenient
dimensionless center of mass energy $\epsilon$ we obtain,
\begin{equation}
  \epsilon=E/E_{12},\qquad \epsilon_{p0}=E_{p0}/E_{12}=1/\zeta.
\label{e:def:epsilon}
\end{equation}

We consider the thermonuclear burning with strong plasma screening in
which both reacting ions are strongly coupled ($\Gamma_1 \gg 1$,
$\Gamma_2 \gg 1$; liquid or solid) but quantum effects are relatively
weak, $\zeta\lesssim 1$. For a one-component ion plasma ($Z_1=Z_2$,
$A_1=A_2$) this would correspond to the range of temperatures $0.34\,
T_p \lesssim T \ll T_l$, where $T_l$ is the temperature of strong
coupling ($\Gamma\simeq 1$), and $T_p=\hbar \omega_p/k_B$ is the ion plasma temperature
determined by the ion plasma frequency $\omega_p=\sqrt{4\pi
Z^2e^2n/m}$ (close to the Debye temperature of the one-component
Coulomb crystal). The temperature range under discussion spans
typically over 1--2 orders of magnitude.

\begin{figure}[t]
\centering
\includegraphics[width=0.45\textwidth]{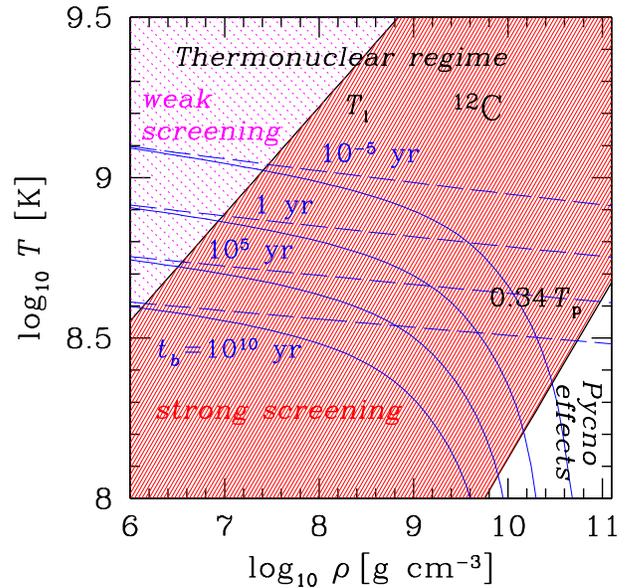}
\caption{(Color online) 
Temperature-density diagram for a $^{12}$C plasma.
Shaded is the temperature--density domain
for
the $^{12}$C matter where carbon burns in the thermonuclear regime.
The domain of dense shading refers to thermonuclear regime with
strong screening. Its upper boundary is determined by the temperature
$T_l$ at which ions become strongly coupled, while the lower boundary
is taken to be $0.34\, T_p$. Four thin solid lines are those at which
characteristic carbon burning time is $t_b=10^{10}$, $10^5$, 1, and
$10^{-5}$ years; four thin dashed lines are the same but neglecting
the plasma physics effects (see text for details).}
\label{fig:diag}
\end{figure}

For example, Fig.\ \ref{fig:diag} is the temperature-density diagram
for the $^{12}$C plasma. The upper thick line is the temperature
$T_l$ of strong Coulomb coupling. Above this line (weakly dashed
region in the upper left corner) carbon is burning in the classical
thermonuclear regime where plasma screening is weak. The lower thick
line is $T =0.34\,T_p$, below which (in the non-dashed region)
pycnonuclear effects in carbon burning become important. The densely
shaded is the domain where the burning is thermonuclear with strong
plasma screening -- the main subject of our study.
To illustrate the efficiency of carbon burning, four thin solid
curves show the lines along which the carbon burning time, defined as
$t_b=n/R$, is constant (from bottom to top), $t_b=$ 10$^{10}$,
$10^5$, 1, $10^{-5}$ years. The reaction rate $R$ is calculated using
the formalism of Ref.~\cite{leandro05}. The plasma physics effects
[of strong screening and pycnonuclear
(``pycno'')
burning]
are included and cause
the bend of the $t_b$ curves at high densities and low temperatures.
Carbon burning is extremely slow below the $t_b=10^{10}$ year line
and
fast above the $t_b=10^{-5}$ year line. Four thin dashed
lines are the same as the thin solid lines but
neglect
the plasma
physics effects. One can see that at high densities and not too high
temperatures these effects are most important. More information on
the the efficiency of carbon burning in the different regimes can be
found, for instance, in Ref.\ \cite{leandro05}.

\section{Model}
\label{s:model}

\begin{figure}[t]
\centering
\includegraphics[width=0.45\textwidth]{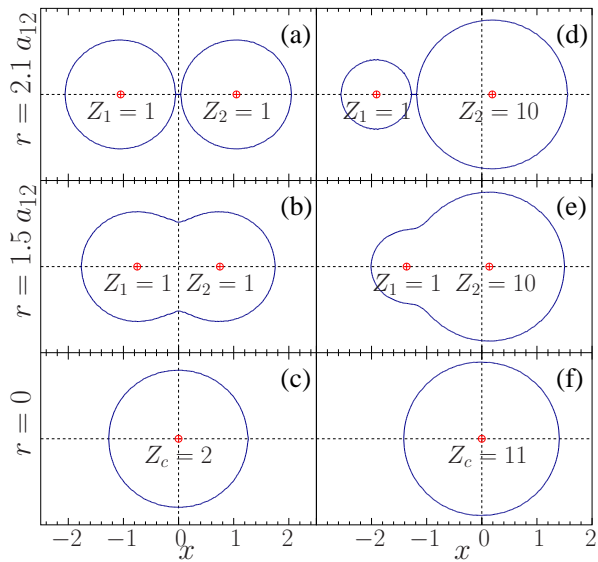}
\caption{(Color online) Simulated shapes of the electron drops around
two colliding nuclei.
(a), (b) and (c): Drop shapes for $Z_2=Z_1$ at
inter-ion distances $2.1\:a_{12}$, $1.5\:a_{12}$ and $0$, respectively.
(d)--(f): Same for  $Z_2=10\,Z_1$. }
\label{fig:shape}
\end{figure}

Let us formulate our model for thermonuclear reactions with strong
plasma screening.

At the first stage (Sec.\ \ref{s:poten}) we introduce the Coulomb
potential $U_C(r)$ for point-like colliding atomic nuclei in the
standard form:
\begin{equation}
    U_C(r)=\frac{Z_1 Z_2 e^2}{r}-H(r),
\label{e:screenpot}
\end{equation}
where $H(r)$ is the mean-field plasma screening potential to be
determined. We will calculate $H(r)$ in the spirit of ion-sphere
model suggested by Salpeter \cite{Salp}. We assume that $H(r)$ is
produced by an electron cloud around the reactants. Such systems
of ions surrounded by electron clouds in dense matter are sometimes called
Onsager molecules (e.g., Ref.\ \cite{rc97} and references therein).
We assume rigid electron charge density,
and view the electron cloud as an incompressible uniformly
charged liquid drop. The electron drop has constant volume; its
charge fully compensates the electric charge of the reactants.
However, it has variable shape which minimizes Coulomb energy of the
system (of the electron drop and the two reactants). It acts as a
Wigner-Seitz cell in which the ions tunnel. In Fig.~\ref{fig:shape}
we show the calculated electron drop shapes for several inter-ion
distances and charge ratios.

This model for $H(r)$ is expected to be adequate in the regime of
strong Coulomb coupling. For a weak coupling, the plasma screening
would be too weak to be described by an electron drop with a sharp
boundary; anyway, it would have no strong effect on thermonuclear
burning rates -- see, e.g., Ref.\ \cite{Salp}.
%
Our model, as
the Salpeter model \cite{Salp}, is so simple
that it does not distinguish the cases of strong coupling in Coulomb
liquid and solid (the effects of background ions are tacitly
described by the electron drop).

At the next stage (Sec.\ \ref{s:sfact}) we calculate the
astrophysical $S$ factors for thermonuclear reactions adding the
screening potential to the total potential in which the nuclei fuse.
In this way we include the screening effect into the $S$ factor, so
that the
modified
$S$ factor becomes determined not only by nuclear
interactions but also by the parameters of dense matter.

Finally, in Sec.\ \ref{s:rate} we calculate the reaction rates $R$
for thermonuclear reactions with the modified astrophysical $S$
factors in a standard way, assuming the Maxwellian velocity
distribution of the reactants. With these rates we determine the
plasma screening enhancement factors $f$ from Eq.\ (\ref{e:enhanc}).

Our model is well defined and easily realized. It will be compared
with other available models for thermonuclear reactions with strong
plasma screening.

\section{Plasma screening potential}
\label{s:poten}

Let us calculate the screening potential $H(r)$ in the electron drop
model. At large separations we have
\begin{equation}
    H(r)=\frac{Z_1 Z_2e^2}{r},
    \quad U_C(r)\equiv 0~~{\rm at}~~r\geq (a_1+a_2).
\label{e:large-r}
\end{equation}
In this case each reacting ion is surrounded by its own ion-sphere of
radius $a_1$ or $a_2$. The electrons within these ion-spheres fully
compensate the ion charges, making the ion spheres electrically
neutral (and, hence, non-interacting).  The electrostatic energy of
these two spheres is
\begin{equation}
    W_{12}=W(Z_1)+W(Z_2)=
    -\frac{0.9e^2}{a_e}\,(Z_1^{5/3}+Z_2^{5/3}).
\label{e:U12}
\end{equation}

At smaller separations, $r<(a_1+a_2)$, the two ion spheres merge,
forming one common electron drop, so that $U_C(r)$ becomes finite.

It is convenient to write,
\begin{equation}
     H(r)=E_{12} h(x),\quad x={r\over a_{12}},
\label{e:defh(x)}
\end{equation}
where $h(x)$ is a dimensionless function of a dimensionless radial
coordinate $x$. The critical separation $r=(a_1+a_2)$ corresponds to
$x=2$, and at $x\geq 2$ we have $h(x)=1/x$. At $x \ll 2$ the function
$h(x)$ is expandable as
\begin{equation}
     h(x)=b_0+b_2x^2+b_4x^4+\dots
\label{e:lowxexp}
\end{equation}
%
The expansion coefficients $b_0$, $b_2$, $b_4, \ldots$ appear to depend on
the only one parameter
\begin{equation}
     z={Z_2}/{Z_1},
\label{e:defz}
\end{equation}
with $z=1$ for equal charges $Z_1=Z_2$, and $z\neq1$ for $Z_1 \neq
Z_2$. The normalized potential $h(x)$ is symmetric with respect to
$z\rightarrow1/z$, so that it is sufficient to consider the case of
$z\geq1$.

The first expansion coefficients are (Appendix~\ref{sec:pert}):
\begin{eqnarray}
    b_0&=&\frac{0.9}{2z}\,\left[(1+z)^{5/3}-1-z^{5/3}\right] \,
    \left( 1+z^{1/3}  \right),
\label{e:b0}\\
    b_2&=&-\frac{1}{16}\,\frac{(1+z^{1/3})^3}{1+z},
\label{e:b2}\\
    b_4&=& \frac{z}{64}\,\frac{(1+z^{1/3})^5}{(1+z)^{11/3}}.
\label{e:b4}
\end{eqnarray}
The expression for $b_0$ was obtained by Salpeter \cite{Salp}; $b_2$
for $z=1$ was derived by Jancovici \cite{jancovici77}, and
generalized for $z\neq1$ by Ogata~\emph{et al.} \cite{oii91}. The
expression for $b_4$ seems original. For $z=1$ we have
$b_0\approx1.0573$, $b_2=-0.25$, $b_4\approx 0.0394$.

Although Eqs.\ (\ref{e:b0})--(\ref{e:b4}) are derived within the
electron-drop model, they accurately describe the real potential
$h(r)$. The applicability of Eq.\ (\ref{e:b0}) has been confirmed by
numerous Monte Carlo (MC) simulations. The coefficient $b_2$ is
basically the contribution from the electron background. As shown by
Jancovici \cite{jancovici77}, neighboring ions do not contribute to
this order, making Eq.~(\ref{e:b2}) quite robust.

The leading term in the expansion (\ref{e:lowxexp}) gives \cite{Salp}
\begin{equation}
  H(0)=H_0=E_{12}\,b_0=W(Z_c)-W_{12},
\label{e:defH_0}
\end{equation}
which is the difference of electrostatic ion-sphere energies for the
compound nucleus and the two reacting nuclei.

It is important that $H(0)$ is accurately determined from numerous
MC simulations of strongly coupled multicomponent ion mixtures. These
simulations are superior to the electron drop model. The results
indicate that strongly coupled mixtures  obey linear mixing rule
(see, e.g., Ref.\ \cite{hpy07}) according to which
\begin{equation}
    H(0)_\mathrm{MC}/k_B T=f_C(\Gamma_1)+f_C(\Gamma_2)-f_C(\Gamma_c),
\label{e:H_MC}
\end{equation}
where $f_C(\Gamma)$ is the Coulomb free energy per one ion in units
of $k_BT$ in a strongly coupled one component plasma. Then the MC
value of $b_0$ is
\begin{equation}
    b_0^\mathrm{MC}=(f_C(\Gamma_1)+f_C(\Gamma_2)-f_C(\Gamma_c))
    /\Gamma_{12}.
\label{e:b0_MC}
\end{equation}
The function $f_C(\Gamma)$ has been accurately calculated by varous
methods and fitted by analytic expressions.
For instance, one can use a fit  from Ref.\ \cite{pc00} (for
one-component ion gas and liquid),
\begin{eqnarray}
   f_C(\Gamma)&=&A_1\,
   \left[\sqrt{\Gamma (A_2+\Gamma)}
   -A_2\ln \left(\sqrt{\Gamma\over A_2}+\sqrt{1+{\Gamma \over A_2}} \right) \right]
\nonumber \\
   &+& 2A_3 \left( \sqrt{\Gamma}- \mathrm{arctan} \sqrt{\Gamma} \right)
\nonumber \\
   &+& B_1 \left[ \Gamma - B_2 \ln \left( 1+ {\Gamma \over B_2}  \right)  \right]
\nonumber \\
   &+& \frac{B_3}{2}\,\ln \left( 1+ \frac{\Gamma^2}{B_4}   \right),
\label{e:f_C}
\end{eqnarray}
where $A_1=-0.907$, $A_2=0.62954$, $B_1=0.00456$, $B_2=211.6$,
$B_3=-0.0001$, $B_4=0.00462$, and
$A_3=-\sqrt{3}/2-A_1/\sqrt{A_2}=0.2771$.

Now let us return to the electron drop model. In addition to the
analytic small-$x$ expansion (\ref{e:lowxexp}), we have calculated
$h(x)$ numerically. The numerical algorithm is as follows. First, two
point-like ions are set at a given separation,
surrounded by
an electron liquid drop of compensating charge symmetrical with
respect to the inter-ion axis. Then several thousands of passes are
run. At each pass we calculate the electrostatic potential and
optimize the drop shape by rearranging small portions of the electron
liquid.

Calculations have been done for the values of $z$ ranging from $1$ to
$10$ with the step of $0.5$. Numerical errors have been estimated by
comparing with the exact analytic results at $r=0$ and $r=a_1+a_2$.
We have also compared the numerical results at low $r$ with the
analytic expansion (\ref{e:lowxexp}) [including the three terms,
Eqs.\ (\ref{e:b0})--(\ref{e:b4})]. The estimated numerical errors are
$\lesssim 0.2\%$.

For convenience of applications we have approximated the numerical
data ($0 \leq x \leq 2$) by an analytic expression
\begin{equation}
    h(x)=\left[ \left( 1- {x^2 \over 4} \right)^2
    (p_0+p_2x^2+p_4x^4) + x^2 \right]^{-1/2},
\label{e:hfit}
\end{equation}
with
\begin{eqnarray}
   p_0 & = & \frac{1}{b_0^2},
\nonumber   \\
   p_2 & = & {p_0 \over 2}-{2b_2 \over b_0^3} -1,
\nonumber \\
   p_4 & = & { 3 b_2^2 \over b_0^4} -
   {2 b_4 \over b_0^3}-{p_0 \over 16}+{p_2 \over 2},
\label{e:pfit}
\end{eqnarray}
$b_0$, $b_2$
and $b_4$ being given by Eqs.\ (\ref{e:b0})--(\ref{e:b4}).
For $z=1$ we have $p_0=0.8945$, $p_2=-0.1297$, and $p_4= -0.0374$.
The analytic fit is constructed in such a way that it correctly
reproduces the small-$x$ expansion (\ref{e:lowxexp}) (including three
terms); it also reproduces the correct value $h(2)=1/2$, smoothly
matching $1/x$ at $x=2$. The formal fit errors are less than 0.2\%
(comparable with the numerical errors).

We have also considered some other $h(x)$ approximations. In
particular, we have tried the polynomial expansion (\ref{e:lowxexp})
in $x^2$ keeping the five terms. The coefficients $b_0$, $b_2$ and
$b_4$ have been set equal to their exact analytic values, while $b_6$
and $b_8$ have been chosen in such a way to match smoothly $h(x)=1/x$
at $x\geq 2$. This approximation is as accurate as (\ref{e:hfit});
the expressions for $b_6$ and $b_8$ are cumbersome (not presented
here).

\begin{figure}[t]
\centering
\includegraphics[width=0.45\textwidth]{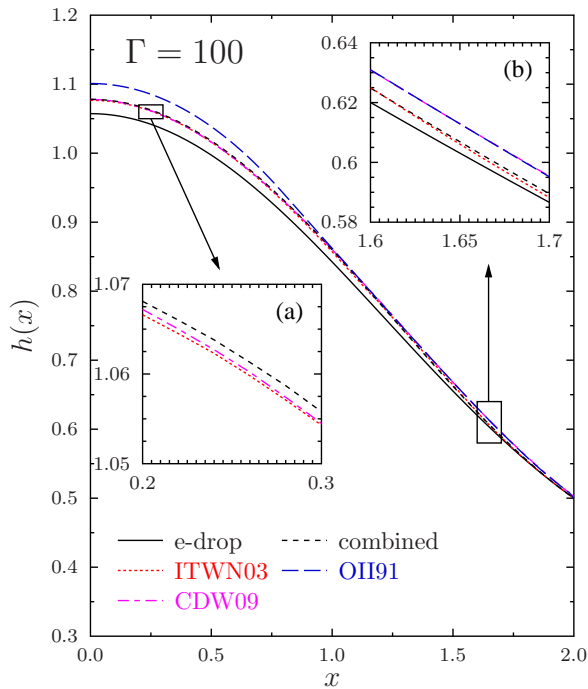}
\caption{(Color online) Screening potential $h(x)$ versus $x$ at
$z=1$. Solid lines show our electron drop (e-drop) fit;  dot-dashed
lines are the fits by Chugunov and DeWitt \cite{Chugunov09};
long-dashed lines are from Ogata~\emph{et al.}\ \cite{oii91},
dotted lines are from Itoh~\emph{et al.} \cite{itwn03};
short-dashed line represents the combined fit. Inserts are zooms
which show the behavior of $h(x)$
at $x\simeq 0.25$ (a), and at
larger $x\simeq 1.65$ (b). See text for details.} \label{fig:pots}
\end{figure}

The screening potential has been studied and approximated by a number
of authors (e.g.\ \cite{oii91,ikm90,itwn03,Chugunov09,caill03} and
references therein). Itoh~\etal~\cite{ikm90} used a not very accurate
approximation matching the $1/x$ behavior with $h(0)$ by linear
functions. Ogata~\etal~\cite{oii91} (OII91) determined the potential
from MC simulations. Their $h(0)$ was improved later
in MC
calculations by Caillol and Gilles \cite{caill03}, as well as by
DeWitt and Slattery \cite{ds99}, and Caillol \cite{caill99}.
Itoh~\etal~\cite{itwn03} (ITWN03) used $h(0)$ obtained
by Jancovici \cite{jancovici77};
relying on the expression
\eqref{e:b2} for $b_2$
they matched the linear behaviour of the potential
 near $x=2$. Chugunov, DeWitt and Yakovlev
\cite{Chugunov07,Chugunov09} (CDW09) constructed the screening
potential from the results of
Ref.~\cite{pc00} and from their own MC data.

In Fig.~\ref{fig:pots} we compare our electron drop results with
those obtained previously (for one-component strongly coupled plasma
of ions). At $x>1$ all curves are in good agreement but slightly
differ from our electron drop fit. At smaller $x$ the OII91 data
deviate from others (due to poor MC statistics in OII91 at small
$x$). Our results differ due to simplicity of the electron drop
model. The combined curve, obtained from our fit \eqref{e:hfit} with
$b_0$ given by Eq.~\eqref{e:b0_MC} instead of the
electron drop value, Eq.~\eqref{e:b0}, closely reproduces the data of
CDW09 and ITWN03. The combined approximation is also discussed
in Sec.\ \ref{s:discussion}.
From the insert (b) we see that
MC-based results (CDW09, OII91) deviate from others. While these results
for the mean-field potential are superior to other fits, this
discrepancy only happens at distances $x \gtrsim 1.2$, whereas for the discussed
case of $\zeta\lesssim 1$ all turning points at Gamow peak energies have
$x<1$.
Note that for $\zeta\gtrsim 1$ our consideration
of thermonuclear burning becomes questionable (Secs.\ \ref{s:Salpeter} and \ref{s:sfact}).

\section{Salpeter's model}
\label{s:Salpeter}

Before we focus on the full electron drop model we outline a simpler
(basic) model for plasma screening in thermonuclear reactions. We
will do it in the spirit of Salpeter's model \cite{Salp} of ion
spheres and call it the Salpeter's model. The quantities calculated
within this model will be labeled by the index $S$. In this model,
the screening potential is replaced by the constant potential $H_0$,
that is given by Eq.~(\ref{e:defH_0}) and corresponds to the leading
term $h(x)=b_0$ in the expansion (\ref{e:lowxexp}). Then the Coulomb
energy (\ref{e:screenpot}) becomes
\begin{equation}
    U_C^{(S)}(r)=\frac{Z_1 Z_2 e^2}{r}-H_0.
\label{e:H0approx}
\end{equation}
This constant ($r$-independent) screening potential $H_0$ is
determined by the density of the matter. For a reaction between
identical nuclei, $H_0=1.0573 Z^2e^2/a_1$.
The respective plasma screening
does not change the shape of $U_C(r)$ but simply lowers the pure
Coulomb potential $Z_1 Z_2 e^2/r$ by $H_0$ which enhances naturally
the nuclear fusion rate.

The well known expression for a thermonuclear reaction rate [s$^{-1}$
cm$^{-3}$] adopted here is
\begin{equation}
 R=\chi\,n_1n_2 I,\quad I=\int_0^\infty dE\,S(E)\,
 \exp\left(-2\pi \eta -\frac{E}{k_B T}\right),
\label{e:rate}
\end{equation}
where $\chi$ is a symmetry factor ($\chi={1 \over 2}$ for a reaction
with identical nuclei, and $\chi=1$ otherwise), $\eta=Z_1 Z_2
e^2/\hbar v$ is the Sommerfeld parameter, $v=\sqrt{2E/\mu}$ is the
relative collision velocity (with kinetic center of mass energy $E$)
at large separations. The factor $-2 \pi \eta$ in the exponent
argument comes from the definition of the astrophysical $S$ factor
(determines the penetration through the pure Coulomb barrier
$U_C^{(0)}(r)=Z_1Z_2e^2/r$), and the factor $-E/k_BT$ comes from the
Maxwellian distribution of reactants over $E$).

In the absence of plasma screening we have $R=R_0=\chi n_1n_1 I_0$,
where the normalized reaction rate $I=I_0$ is
\begin{equation}
  I_0=\int_0^\infty dE\,S_0(E)\,
 \exp\left(-2\pi \eta -\frac{E}{k_BT}\right),
\label{e:rate0}
\end{equation}
and $S_0(E)$ is the standard astrophysical $S$ factor calculated
without any screening. In many cases the integral over $E$ can be
taken quite accurately using the saddle-point method, and this
well-known result is
\begin{equation}
 I_0=4 \, \sqrt{\frac{2E_{p0}}{3 \mu}}
 \,\frac{S_0(E_{p0})}{k_BT} \exp(-\tau),
\label{e:R0}
\end{equation}
where $E_{p0}=k_BT\tau/3$ is the Gamow peak energy, and $\tau$ is
given by Eq.~(\ref{tau12}).

Now we calculate the Salpeter's rate, $R=R_S=\chi n_1 n_2 I_S$,
including plasma screening under the following simplified
assumptions:
\begin{enumerate}
\item The screening potential is given by
(\ref{e:H0approx});

\item Coulomb barrier is thick; the barrier penetration is
calculated in the WKB approximation;

\item The barrier penetration is the same as
in $s$ wave ($\ell=0$).

\end{enumerate}

The normalized reaction rate $I_S$ is then given by the same equation
(\ref{e:rate0}) as $I_0$ but with $S_0(E)$ replaced by $S_S(E)$ to
account for the plasma screening in the Salpeter's model. The
function $S_S(E)$ is determined by the quantum barrier penetration
times $\exp(2\pi\eta)$, and the WKB penetration factor is (see Sec.\
\ref{s:sfact})
\begin{equation}
  \exp\left(-\frac{2}{\hbar}\,\int_{r_1}^{r_2} dr\,
  \sqrt{2 \mu (U_\mathrm{eff}(r)-E)}
  \right) .
\label{e:factor0}
\end{equation}
In this case $U_\mathrm{eff}(r)$ is the $s$-wave effective potential
which includes the nuclear and Coulomb components, while $r_1$ and
$r_2$ are the classical inner and outer barrier penetration (turning)
points, respectively. If we include the Salpeter's screening, we have
$U_\mathrm{eff}(r) \to U_\mathrm{eff}(r)-H_0$. Therefore, replacing
$E \to E'=E+H_0$ we keep the expression for the factor
(\ref{e:factor0}) unchanged. With this in mind it is easy to show
that
\begin{equation}
 I_S  =  \int_0^\infty dE\,S_S(E)\,
 \exp\left(-2\pi \eta -\frac{E}{k_BT}\right)
 = f_S' I_S',
 \label{e:I1}
\end{equation}
with
\begin{equation}
  I_S'= \int_{H_0}^\infty dE'\,S_0(E')\,
 \exp\left(-2\pi \eta(E') -\frac{E'}{kT}\right),
\label{e:I11}
\end{equation}
and
\begin{equation}
  f_S'=\exp \left(  \frac{H_0}{k_BT}   \right).
\end{equation}
In Eq.~(\ref{e:I1}) we have introduced
\begin{equation}
    S_S(E)=S_0(E+H_0)\,\exp(2\pi\eta(E)-2\pi\eta(E+H_0)),
\label{e:S1}
\end{equation}
which can be called the Salpeter's $S$ factor corrected for plasma
screening effects (within the Salpeter's model).
Notice that Salpeter did not directly include plasma screening into
$S(E)$ but his calculations can be treated in this way.

Now the enhancement factor $f_S$ of nuclear reactions in the
Salpeter's model becomes
\begin{equation}
   f_S=\frac{R_S}{R_0}=f_S'\,\frac{I_S'}{I_0}.
\end{equation}
If the integral (\ref{e:I11}) can be calculated by the standard
saddle-point method and the integrand function has a traditional
Gamow-peak shape with the Gamow-peak window above $H_0$, we can shift
the lower integration limit to 0 and immediately obtain $I_S'=I_0$.
Then the enhancement factor acquires the standard Salpeter's form
\cite{Salp}:
\begin{equation}
   f_S=f_S'=\exp \left(  \frac{H_0}{k_BT}   \right)=\exp (\Gamma_{12}b_0).
\end{equation}
In the regime of strong Coulomb coupling this factor can be huge.

As a byproduct of the saddle-point integration in Eq.~(\ref{e:I11})
we obtain that the Gamow peak energy in the Salpeter's model is
$E_{pS}'=E_{pS}+H_0=E_{p0}$, and the outer turning point is
unaffected by the screening. Then
\begin{equation}
  E_{pS}={\tau k_BT \over 3}-H_0, \quad
  r_2=\frac{3Z_1Z_2 e^2}{k_BT\tau}.
\label{e:Epkr2Salp}
\end{equation}
Therefore, when the temperature decreases, $E_{pS}$ goes down and
(formally) can become negative. The Gamow-peak window moves then out
of the integration region $E>0$ in Eq.~(\ref{e:I1}) and the
saddle-pint method becomes inapplicable. The condition $E_{pS}=0$
corresponds to $\zeta\approx 1$, with $\zeta$ defined by
Eq.~(\ref{tau12}). At these low temperatures the thermonuclear
reaction regime breaks down and the present formalism becomes
questionable. Accordingly, in Fig.\ \ref{fig:diag} we restrict the
domain of thermonuclear burning by the $\zeta=1$ line ($T\approx
0.34\, T_p$, where  the ion plasma temperature $T_p$ is defined in
Sec.\ \ref{s:param}).

\begin{figure}[t]
\centering
\includegraphics[width=0.45\textwidth]{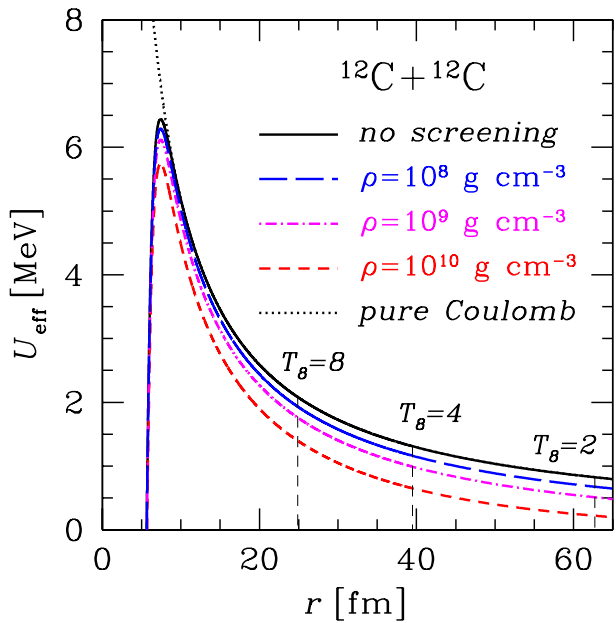}
\caption{(Color online) Effective potential for the $^{12}$C+$^{12}$C
reaction in carbon matter. The solid line is the potential neglecting
plasma screening effects. The long-dashed, dot-dashed, and
short-dashed lines are the potentials reduced by plasma screening in
the regime of strong Coulomb coupling at $\rho=10^8$, $10^9$, and
$10^{10}$ g~cm$^{-3}$, respectively. The dotted line is the pure
Coulomb potential for point-like nuclei. Three thin vertical lines
position the outer turning point $r_2$, Eq.~(\ref{e:Epkr2Salp}), for
barrier penetration with the Gamow-peak energy $E_{pS}$ in the
Salpeter's model at (from left to right) $T_8=T/10^8$ K=8, 4, and 2,
respectively. See text for details.} \label{fig:ccpot}
\end{figure}

For illustration, Fig.\ \ref{fig:ccpot} displays the effective
potential $U_\mathrm{eff}(r)$ for the $^{12}$C+$^{12}$C reaction in
dense carbon matter (in the $\ell=0$ channel) including the nuclear
and Coulomb potentials. The solid line is the standard theoretical
potential which is taken from Ref.\ \cite{yak10} and neglects plasma
screening effects. It is almost pure Coulomb at $r \gtrsim 9$ fm but
it is strongly dominated by nuclear attraction at lower $r$. The
three lower lines (long-dashed, dashed-dot, and short-dashed) are
obtained taking into account the plasma screening in the regime of
strong Coulomb coupling (the dashed domain in Fig.\ \ref{fig:diag})
at $\rho=10^8$, $10^9$, and $10^{10}$ g~cm$^{-3}$. The Coulomb part
of $U_\mathrm{eff}(r)$ is corrected for the plasma screening. For the
displayed conditions, the Salpeter's Eq.~(\ref{e:H0approx}) is an
excellent approximation for $U_C(r)$. At $\rho=10^8$, $10^9$, and
$10^{10}$ g~cm$^{-3}$ the plasma screening reduces the potential by
$H_0= 0.151$, 0.326, and 0.702 MeV, respectively. The higher the
density the stronger the screening effect (the larger the enhancement
factor).

Thin vertical lines in Fig.~\ref{fig:ccpot} position the outer
turning points $r_2$,  Eq.~(\ref{e:Epkr2Salp}), for the barrier
penetration with the Gamow peak energy at $T=8 \times 10^8$, $4
\times 10^8$ and $2 \times 10^8$~K ($r_2=$24.88, 39.49 and 62.69 fm,
respectively). The Gamow peak energy for these $T$ is $E_{pS}=$2.08,
1.31 and 0.827 MeV. Figure~\ref{fig:ccpot} clearly shows the ranges
of sub-barrier distances which regulate the barrier penetration in
the WKB approximation. Only these distances are important for the
nuclear reaction problem, and in all the displayed cases the
Salpeter's potential (\ref{e:H0approx}) serves as a very good
approximation. It would be not distinguishable from the exact
screened potential in Fig.\ \ref{fig:ccpot}. At larger $r$, the
Salpeter's potential would be noticeably different from from the
exact one, especially at the highest assumed $\rho=10^{10}$
g~cm$^{-3}$ (in which case the difference would be pronounced at $r
\gtrsim 80$ fm). However, these large $r$ could contribute to the
nuclear reaction rates only at lower $T$ at which our approach
becomes invalid (pycnonuclear effects, which we neglect, would become
strong).

\begin{figure}[t]
\centering
\includegraphics[width=0.45\textwidth]{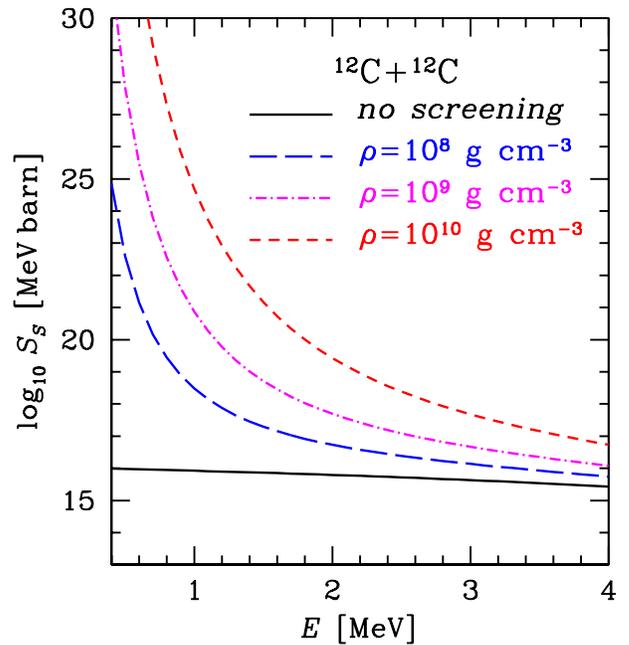}
\caption{(Color online) Astrophysical $S$ factor as a function of the
center of mass energy $E$ of reactants for the $^{12}$C+$^{12}$C
reaction in carbon matter calculated in the Salpeter's model. The
solid line is the standard $S$ factor neglecting plasma screening.
The long-dashed, dot-dashed, and short-dashed lines are the $S$
factors enhanced by plasma screening in the regime of strong Coulomb
coupling at $\rho=10^8$, $10^9$, and $10^{10}$ g~cm$^{-3}$,
respectively.} \label{fig:sfcc}
\end{figure}

\begin{figure}[t]
\centering
\includegraphics[width=0.45\textwidth]{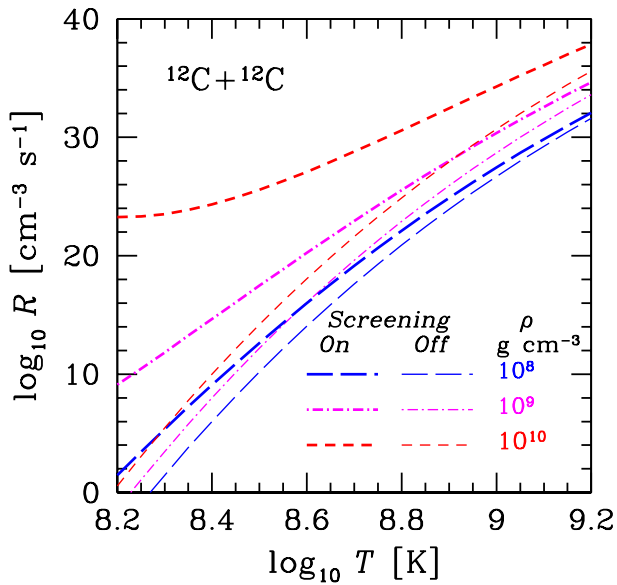}
\caption{(Color online) $^{12}$C+$^{12}$C thermonuclear reaction rate
in carbon matter as a function of temperature. The long-dashed,
dot-dashed, and short-dashed lines correspond to $\rho=10^8$, $10^9$,
and $10^{10}$ g~cm$^{-3}$, respectively. Thin lines are calculated
neglecting the plasma screening, while thick lines include the
screening in the Salpeter's model (see text for details).}
\label{fig:thermorate}
\end{figure}

In Fig.\ \ref{fig:sfcc} we show the astrophysical $S$ factor for the
$^{12}$C+$^{12}$C reaction in carbon matter as calculated in the
Salpeter's model from Eq.\ (\ref{e:S1}). The solid line is the
standard theoretical $S$ factor unaffected by plasma screening. The
long-dashed, dot-dashed, and short-dashed lines are the $S$ factors
calculated with account for plasma screening at $\rho=10^8$, $10^9$,
and $10^{10}$ g~cm$^{-3}$, respectively. The screening effects
increase the transparency of the Coulomb barrier in dense plasma
(Fig.\ \ref{fig:ccpot}). These effects are especially pronounced at
low energies $E$ and high densities $\rho$ where the $S$ factor is
enhanced by many orders of magnitude. Naturally it dramatically
enhances nuclear reaction rates.

Figure \ref{fig:thermorate} presents the $^{12}$C+$^{12}$C reaction
rate in carbon matter as a function of temperature for the same three
values of $\rho$. Again, the long-dashed, dot-dashed, and
short-dashed lines refer to $\rho=10^8$, $10^9$, and $10^{10}$
g~cm$^{-3}$, respectively. As seen from Fig.\ \ref{fig:diag}, the
ranges of $T$ and $\rho$ chosen in Fig.\ \ref{fig:thermorate}
correspond to thermonuclear burning with strong plasma screening. The
thin lines show the rates calculated neglecting the screening. The
thick lines include the screening in the Salpeter's model. The
screening always enhances the rate but does not prevent the decrease
of the rate with the fall of $T$ as long as the burning regime is
thermonuclear (the rate becomes temperature-independent only in the
pycnonuclear regime). It is seen that the screening enhancement of
thermonuclear reaction rate increases at lower $T$. For $\rho=10^8$
g~cm$^{-3}$ the screening effect is not very strong but at
$\rho=10^{10}$ g~cm$^{-3}$ and lowest displayed $T\approx 1.5 \times
10^8$~K it is tremendous: it enhances the reaction rate by more than
20 orders of magnitude (also see Sec.\ \ref{s:discussion}).

\section{Astrophysical $S$ factor}
\label{s:sfact}

\subsection{Screening effects on $S$ factor}

At the next step we use the liquid drop model and calculate
astrophysical $S$ factors and thermonuclear reaction rates beyond the
Salpeter's model.

In the barrier penetration model the astrophysical $S$ factor is
given by
\begin{equation}
    S(E) = \exp(2\pi\eta)\,\frac{\pi\hbar^2}{2\mu}
    \sum_{\ell} (2\ell+1)T_\ell(E)P_\ell(E), \label{eq:sdef}
\end{equation}
where $\eta$ is defined in Sec.\ \ref{s:Salpeter}, $T_\ell$ is the
barrier penetration probability, $P_\ell$ is the nuclear fusion probability,
and the sum is over angular momenta $\ell$. For low energies of
astrophysical interest it is usually sufficient to set $P_\ell(E)=1$.

At subbarrier energies the barrier penetration probability is small
and can be described by the WKB formula
\begin{align}
    T_\ell(E)&=\exp\left[-s_\ell(E)\right],\nonumber \\
    s_\ell(E)&=\frac{2}{\hbar} \int_{r_1}^{r_2}dr\,
    \sqrt{2\mu\left(U_\mathrm{eff}(r)+\frac{\hbar^2 \ell(\ell+1)}
    {2\mu r^2}-E\right)}, \label{eq:tl}
\end{align}
where $U_\mathrm{eff}(r)$ is the effective potential at $\ell=0$
(Fig.\ \ref{fig:ccpot}). The potential contains the standard
potential, that neglects plasma screening, minus $H(r)$ given by
Eqs.~(\ref{e:defh(x)}) and (\ref{e:hfit}).

As discussed in Sec.~\ref{s:Salpeter}, one of the screening effects
on $S(E)$ would be to shift $E \to E+H_0$, which corresponds to
\begin{equation}
     S(E)\to S_0(E'),\quad E'=E+H_0.
\label{e:Sf}
\end{equation}
Therefore, we can write
\begin{equation}
    S(E)=S_0(E')q(E),
\end{equation}
where $q(E)$ is an extra correction factor to be analyzed.

Introducing dimensionless units (Sec.\ \ref{s:param}) we rewrite
Eq.~\eqref{eq:tl} as
\begin{align}
     s_\ell(\epsilon)=\frac{4\Gamma_{12}}{\pi\zeta^{3/2}}
     \int_{x_1}^{x_2}dx\,
     \sqrt{v(x)+\frac{\pi^2\zeta^3\,\ell(\ell+1)}
     {4\Gamma_{12}^2\,x^2}-\epsilon},
\label{tldef}
\end{align}
where $v(x)=U_\mathrm{eff}(r)/E_{12}$, $\epsilon=E/E_{12}$,
$x=r/a_{12}$, with $v(x)=0$ and $h(x)=1/x$ for $x\geq2$.

According to Eq.~(\ref{eq:sdef}) we can write
\begin{equation}
    q(\epsilon)=q_0(\epsilon) q_L(\epsilon),
\label{eq:Rmain}
\end{equation}
where $q_0(\epsilon)$ and $q_L(\epsilon)$ include the corrections due
to $\ell=0$ and $\ell>0$ channels, respectively. Using \eqref{eq:tl}, we have
\begin{equation}
   q_0(\epsilon)=\exp\left(2\pi\eta-2\pi\eta'-
   s_0+s_{00}'\right),
\label{e:R0}
\end{equation}
and
\begin{equation}
    q_L(\epsilon)= \frac{\sum_\ell (2\ell+1) \exp\left(
    s_0-s_\ell\right)}{\sum_\ell (2\ell+1)
    \exp\left(s_{00}'-s_{0\ell}'\right)}, \label{e:RL}
\end{equation}
where summation is over all $\ell \geq 0$, $\eta=\eta(\epsilon)$,
$\eta'=\eta(\epsilon')$, $s_\ell=s_\ell(\epsilon)$,
$s_{\ell}'=s_\ell(\epsilon')$, with $s_{00}'$ and $s_{0\ell}'$ being
the quantities calculated neglecting screening.

Evidently, the $S$ factors are affected by nuclear physics and plasma
physics effects. In the thermonuclear regime these effects are
usually decomposed and studied separately. We will mainly focus on
the plasma screening effects, which are determined by differences
like $s_{00}'-s_0$. They are expressed as integrals whose integrands
are small at those conditions at which nuclear forces are
significant. To analyze these integrals, it is sufficient to neglect
the nuclear potential in $U_\mathrm{eff}(r)$ in the expressions for
$q(E)$, assuming thus $U_\mathrm{eff}(r)=U_C(r)$. In this
approximation the inner turning point goes to zero, $r_1\to 0$ ($x_1
\to 0)$.

\subsection{Screening correction in $s$-wave}
\label{sec:leqn}

Here we study the factor $q_0$ given by Eq.~\eqref{e:R0}. We have
\begin{equation}
    s_0-s_{00}'={\Gamma_{12}}{\zeta^{-3/2}}
    \left[F(\epsilon)-F_0(\epsilon')\right],\label{eq:approx}
\end{equation}
where
\begin{align}
    &F_0(\epsilon)=\frac{4}{\pi}\int_0^{x_{02}}dx\,\sqrt{{1\over x}-\epsilon}
    =\frac{2}{\sqrt{\epsilon}}, \label{eq:defFf}\\
    &F(\epsilon)=\frac{4}{\pi}\int_0^{x_2}dx\,G_\epsilon(x),
    \quad G_\epsilon(x)=\sqrt{{1\over x}-h(x)-\epsilon}.
\label{eq:defFs}
\end{align}
The term $\Gamma_{12}\zeta^{-3/2}F_0(\epsilon')$ cancels $-2\pi\eta'$ in
Eq.~(\ref{e:R0}), so that
\begin{equation}
    q_0(\epsilon)=\exp\left(-{\Gamma_{12}}{\zeta^{-3/2}}
    \left[F(\epsilon)-F_0(\epsilon)\right]\right).
\label{eq:Dresult}
\end{equation}

The higher the energy $\epsilon$ (or $\epsilon'$), the lower the
turning point $x_2$ in~Eq.~\eqref{eq:defFs}. Therefore, we can derive the
high-$\epsilon$ expansion corresponding to the small-$x$
expansion~\eqref{e:lowxexp}. For this purpose we rewrite the
integral~\eqref{eq:defFs} in the form
\begin{equation}
  F(\epsilon)
  =\frac{4}{\pi}\int_0^{\infty} x_2(G^2+\epsilon')\,dG, \label{e:FinG}
\end{equation}
where $x_2(\lambda)$ is the solution to the equation
$\lambda-\epsilon'=1/x_2-h(x_2)-\epsilon$.
Taking $\lambda=\epsilon'$ we see that
$x_2(\epsilon')$ (at $G=0$) is indeed the turning point $x_2$,
justifying the notation. The
equation for $x_2(\lambda)$ can
be rewritten as
\begin{equation}
  x_2=\frac{1-b_2x_2^3-b_4x_2^5-...}{\lambda},
\end{equation}
which is easily iterated to obtain the series expansion
\begin{equation}
    x_2(\lambda)=
    \frac{1}{\lambda}-\frac{b_2}
    {\lambda^4}-\frac{b_4}{\lambda^6}+\frac{3b_2^2}{\lambda^7}
    +\ldots
\label{e:x2.epsil}
\end{equation}
Substituting it into Eq.~\eqref{e:FinG} and integrating we obtain
\begin{align}
    F(\epsilon)=&F_0(\epsilon')\left(1-\frac{5b_2}{16\,\epsilon'^{3}}
    -\frac{63b_4}{256\,\epsilon'^{5}}
    +\frac{693 b_2^2}{1024\,\epsilon'^{6}}+\ldots \right).
\label{eq:analyticF}
\end{align}

We have not rigorously examined the convergence of this series, but
it seems asymptotic.
In any case, for the low energies of interest, the higher order terms
do not considerably increase the accuracy. Thus we can safely omit
the last term.

To have an accurate expression for $F(\epsilon)$ we also have
calculated $F(\epsilon)$ numerically and fitted the results by
\begin{equation}
    F(\epsilon)=F_0(\epsilon')\left(1-\frac{5b_2}{16\,\epsilon'^{3}}
    -c_1\,\frac{63b_4}{256\,\epsilon'^{5}}
    +c_2\,\frac{693 b_2^2}{1024\,\epsilon'^{6}} \right),
    \label{e:F.epsilon}
\end{equation}
where $c_1=3.662$ and $c_2=2.762$ are two fit parameters. The ranges
of $\epsilon$ and $z$ employed in the fit are $0.00025 \lesssim
\epsilon\lesssim 9$ and $1\leq z\leq 10$. The maximum relative fit
error is $1.6\,\%$ (at $z=1$ and $\epsilon\approx0.00025$) and the
absolute rms error is $10^{-4}$. For $0.05\leq\epsilon\leq2.0$ and
the same $z$ the maximum relative error is $0.6\,\%$. Nevertheless,
we will not use this fit formula further but restrict ourselves by
the analytic expansion (\ref{eq:analyticF}).

Note that Eq.~(\ref{eq:Dresult}) with $F(\epsilon)=F_0(\epsilon')$
corresponds to the Salpeter's model, while the bracketed factors in
Eqs.\  (\ref{eq:analyticF}) and (\ref{e:F.epsilon}) are the
corrections provided by the electron drop model.

\subsection{Screening correction due to higher-$\ell$ waves}
\label{sec:lneqn}

Here we analyze the factor $q_L(\epsilon)$ which is given by Eq.\
\eqref{e:RL} and provides the screening correction to $S(E)$ due to
$\ell \geq 1$ reaction channels.

We expand~\eqref{tldef} in powers of centrifugal energy keeping the
first-order term:
\begin{equation}
    s_0(\epsilon)-s_\ell(\epsilon)=-\frac{\pi\zeta^{3/2}\ell(\ell+1)}
    {\Gamma_{12}}\,g(\epsilon),
\label{e:S.diffs}
\end{equation}
where
\begin{equation}
    g(\epsilon)=\int_{x_1}^{x_2}
    \frac{dx}{2x^2 \sqrt{v(x)-\epsilon}}.
\label{e:S.defg}
\end{equation}
Higher-order terms in (\ref{e:S.diffs}) are suppressed by extra
powers of $\Gamma_{12}^{-1}$.  In this approximation the turning
points $x_1$ and $x_2$ in $g(\epsilon)$ are the same as for $s$-wave.
Now we treat the exponent arguments in \eqref{e:RL} as small and
replace the sums by integrals. It turns out to be a good
approximation for the ratio (but poorer approximation for the sums
themselves). We obtain
\begin{equation}
    q_L(\epsilon)=\frac{g_0(\epsilon')}{g(\epsilon)}
    =\frac{g_0(\epsilon')}{g_0(\epsilon')+\delta
    g(\epsilon)},
\end{equation}
where $g_0(\epsilon)$ is given by the same Eq.\ (\ref{e:S.defg}) as
$g(\epsilon)$ but neglecting plasma screening, and $\delta
g(\epsilon)=g(\epsilon)-g_0(\epsilon')$. The function $\delta
g(\epsilon)$ can be calculated neglecting nuclear interaction and
taking $x_1\to 0$. In this approximation it is universal --
independent of a particular reaction. The function $g_0(\epsilon)$ is
independent of plasma screening; it can be calculated for a given
reaction using a suitable model of nuclear interaction (for instance,
employing the same formalism as in Refs.\ \cite{yak10,Jacovlev12}).

In analogy to \eqref{eq:analyticF}, we have derived a high-$\epsilon$
expansion for $\delta g$:
\begin{equation}
    \delta g(\epsilon)=-\frac{3\pi b_2}{8\epsilon'^{5/2}}
    -\frac{35\pi b_4}{64\epsilon'^{9/2}}
    +\frac{315\pi b_2^2}{256\epsilon'^{11/2}}+\ldots
\end{equation}

Let us restrict ourselves by $\zeta\leq 1$. For $\zeta=1$ and $z=1$
the Gamow peak energy  is about $\epsilon_p\simeq 0.24$ [see
Eq.~\eqref{e:deltaEpk} below]. In this case the above expansion gives
$\delta g\simeq 0.19$, while the integral \eqref{e:S.defg} gives
$\delta g\simeq 0.21$. For smaller $\zeta$ and higher $\epsilon_p$
the expansion is even more exact.

While $0<\delta g \lesssim 0.2-0.3$, $g_0$ is typically greater than
one. To verify the last statement we have taken the reaction database
from Ref.\ \cite{Jacovlev12}. It contains about 5,000 reactions
involving stable and unstable isotopes of 10 elements (Be, B, C, N,
O, F, Ne, Na, Mg, and Si). We have checked directly that $g_0 \gtrsim
1$ for the reactions between nuclei which are close to the stability
valley at those densities and temperatures where Gamow peak energies
are lower than barrier hight (otherwise reactions are so fast that
the nuclei do not exist in dense matter). Therefore, for many
reactions of practical interest $q_L$ does not deviate from $q_L=1$
more than by $\sim20\%$, which is well within expected uncertainties
in the $S$ factors and reaction rates. To avoid unnecessary
complications we suggest to disregard screening corrections due to
higher-$\ell$ waves and set $q_L=1$.

\section{Plasma screening enhancement of reaction rates}
\label{s:rate}

As seen from Eq.\ (\ref{e:rate}), the plasma screening enhancement
factor of thermonuclear reaction rate is
\begin{equation}
    f=I/I_0,
\label{e:fff}
\end{equation}
where $I$, expressed in our dimensionless units (Sec.\
\ref{s:param}), is given by
\begin{equation}
    I=E_{12} \int_0^\infty S_0(\epsilon')
    \exp\left[-\Gamma_{12}\left(\zeta^{-3/2}
    F(\epsilon)+\epsilon\right)\right]\,d\epsilon.
\label{e:I*}
\end{equation}

Generally, one can calculate $f$ numerically from Eq.~(\ref{e:fff}).
Instead, we analyze $f$ analytically by employing the traditional
Gamow peak formalism. With decreasing temperature, the Gamow peak
becomes narrower which makes the formalism more accurate. On the
other hand, the peak energy decreases, and at $\zeta \approx 1$
(i.e., at $T \approx 0.34 \,T_p$) the peak energy goes to zero (Sec.\
\ref{s:Salpeter}) which manifests the breakdown of the thermonuclear
burning regime and the transition to pycnonuclear burning. We will
restrict ourselves to $\zeta \lesssim 1$; at higher $\zeta$ the
present formalism can be considered as approximation.

In the Gamow peak formalism we can decompose $f$ as
\begin{equation}
   f=f_{scr}f_{Sfact},
\label{e:f-decomposition}
\end{equation}
where $f_{scr}$ is calculated from Eq.~(\ref{e:fff}) neglecting the
dependence of the astrophysical $S$ factor on energy $\epsilon$,
while
\begin{equation}
   f_{Sfact}=S_0(\epsilon'_{p})/S_0(\epsilon_{p0}),
\label{e:fSfact}
\end{equation}
takes into account that plasma screening (beyond the Salpeter model)
shifts the Gamow peak (from $\epsilon_{p0}$ to $\epsilon_p$).

We will mainly focus on $f_{scr}$ and discuss $f_{Sfact}$ briefly in
Sec.\ \ref{s:discussion}. It is traditional to express $f_{scr}$ as
\begin{equation}
    f_{scr}=\exp\left[h_0(\Gamma_{12})+h_1(\Gamma_{12},\zeta)\right].
\label{e:f_scr}
\end{equation}
Here, $h_0(\Gamma_{12})=\Gamma_{12} b_0 =H(0)/(k_BT)$ is the leading
term of the normalized screening potential $h(x)$. This term assumes
quantum tunneling in a Coulomb potential lowered by a constant value
$H(0)$ (equivalent to the Salpeter's approximation, Sec.\
\ref{s:Salpeter}). This term does not change the Coulomb potential
shape, and therefore does not affect the dynamics of quantum
tunneling. The next term $h_1(\Gamma_{12},\zeta)$ in the exponent of
Eq.~(\ref{e:f_scr}) is the correction to $h_0(\Gamma_{12})$; it is
produced by the variation of the mean-field potential shape due to
screening over quantum tunneling path; it is generally smaller than
$h_0(\Gamma_{12})$.

Let us take Eq.\ (\ref{e:I*}) and use the expansion
(\ref{eq:analyticF}) (treating the $1/\epsilon'^{3}$ and
$1/\epsilon'^{5}$ terms in the parentheses as small corrections and
neglecting the $1/\epsilon'^{6}$ term). Furthermore we ignore the
dependence of $S_0$ on $\epsilon$, and take the integral $I$ by the
standard saddle-point method. This gives the dimensionless Gamow peak
energy
\begin{equation}
  \epsilon'_p=\frac{1}{\zeta}+\delta \epsilon_p,\quad
  \delta \epsilon_p=-\frac{35}{24}\,b_2\zeta^2-\frac{231}{128}\,b_4\zeta^4.
\label{e:deltaEpk}
\end{equation}
The leading term $1/\zeta$ corresponds to the Salpeter's
approximation [see Eqs.\ (\ref{e:def:epsilon}) and
(\ref{e:Epkr2Salp})], while $\delta \epsilon_p$ is a small correction
beyond this approximation.

With such a correction the saddle-point method gives
\begin{equation}
    h_1(\Gamma_{12},\zeta)=\Gamma_{12} \left( \frac{5}{8}\,b_2\zeta^2
    +\frac{63}{128}\,b_4\zeta^4\right).
\label{eq:analytich1}
\end{equation}
Eqs.\ (\ref{e:deltaEpk}) and (\ref{eq:analytich1}) represent
truncated expansions in powers of $\zeta$.  Higher-order terms
(starting with $\zeta^5$) can be determined but seem unimportant for
applications. The advantage of these equations is that they are
derived from first principles. They are valid for any mean field
screening potential (not only for the electron drop model). In order
to use them one needs the three coefficients, $b_0$, $b_2$, and
$b_4$, for a given mean field model. Note that the leading term in
$h_1(\Gamma_{12},\zeta)/\Gamma_{12}$ is ${5 \over 8}\,b_2\zeta^2$; it
is well known \cite{aj78}; the second term seems original.

\begin{figure}[t]
\centering
\includegraphics[width=0.45\textwidth]{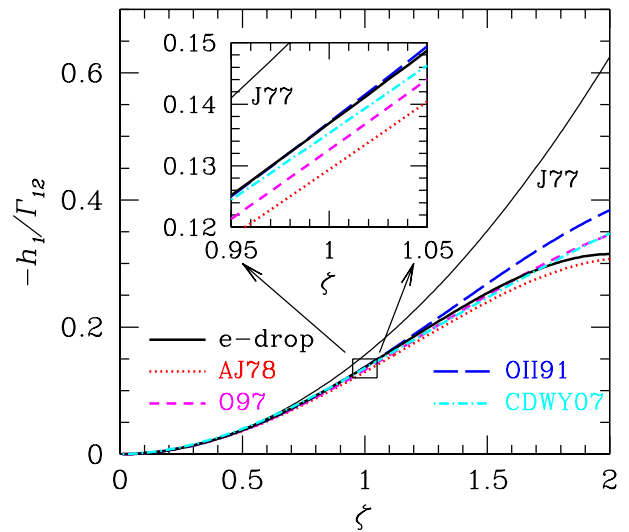}
\caption{(Color online) Function
$-h_1(\Gamma_{12},\zeta)/\Gamma_{12}$ versus $\zeta$ in the regime of
strong Coulomb coupling at $Z_1=Z_2$. The thin solid line J77
corresponds to the leading term of Eq.\ (\ref{eq:analytich1}) derived
by Jancovici \cite{jancovici77}; thick lines of various types are
calculations by different authors (AJ78 -- \cite{aj78}; O97 --
\cite{ogata97}; OII91 -- \cite{oii91}; CDWY07 -- \cite{Chugunov07};
e-drop -- present work). The insert is a zoom of the
behavior of the different curves at $\zeta \approx 1$ (see text for
details).} \label{fig:h1}
\end{figure}

By way of illustration, in Fig.\ \ref{fig:h1} we present the function
$h_1(\Gamma_{12},\zeta)/\Gamma_{12}$ versus $\zeta$ at $Z_1=Z_2$ and
strong Coulomb coupling as calculated by several authors. The insert
zooms in the behavior of the different curves at $\zeta \approx 1$.
The thin solid curve J77 shows the leading term in Eq.\
(\ref{eq:analytich1}); it was derived by Jancovici
\cite{jancovici77}. The thick solid curve (e-drop) is our electron
drop model; the dotted curve AJ78 is derived by Alastuey and
Jancovici \cite{aj78}; the short-dashed curve O97 by Ogata
\cite{ogata97}; the long-dash curve OII91 by Ogata, Iyetomi and
Ichimaru \cite{oii91}; and the dot-dashed curve CDWY07 by Chugunov,
DeWitt and Yakovlev \cite{Chugunov07}. In the electron drop model
$h_1(\Gamma_{12},\zeta)/\Gamma_{12}$ is independent of $\Gamma_{12}$.
For other models it is a slowly varying function of $\Gamma_{12}$. In
these cases, we set $\Gamma_{12}=150$. Although the present formalism
is strictly valid at $\zeta \lesssim 1$, we extend the plot to
$\zeta=2$ to demonstrate the diversity of results by different
authors. These results are usually presented as analytic fits to
numerical calculations by various methods. We see that our analytic
Eq.~(\ref{eq:analytich1}) at $\zeta \lesssim 1$ agrees very well with
other results. The divergency of the results at $\zeta \gtrsim 2$ is
unimportant for us because we restrict ourselves to the thermonuclear
reaction regime at $\zeta < 1$.

\section{Discussion}
\label{s:discussion}

The main practical outcome of our consideration is that the
enhancement factor of a non-resonant fusion reaction due to strong
plasma screening is given by Eqs.\ (\ref{e:f_scr}) and
(\ref{eq:analytich1}). Combined with $h_0(\Gamma_{12})=\Gamma_{12}
b_0$, these equations give
\begin{equation}
  f_{scr}=\exp\left[\Gamma_{12} \left(b_0+ \frac{5}{8}\,b_2\zeta^2
    +\frac{63}{128}\,b_4\zeta^4\right) \right],
\label{e:ffin}
\end{equation}
expressing $f_{scr}$ through the three coefficients ($b_0$, $b_2$ and
$b_4$) which can be slowly varying functions of plasma parameters and
charge numbers $Z_1$, $Z_2$ of the reactants. This formula is derived
from first principles in the mean field approximation; it is expected
to be valid for any mean field model of the screening potential. Its
applicability is restricted by strong screening ($\Gamma_{12}\gtrsim
1$; $T \lesssim T_l$ in Fig.\ \ref{fig:diag}) and thermonuclear
burning regime ($\zeta \lesssim 1$; $T \gtrsim 0.34\,T_p$ in Fig.\
\ref{fig:diag}). The $b_0$ (Salpeter's) term in Eq.\ (\ref{e:ffin})
is leading while other terms are relatively less important.

We can point out two mean-field models of the screening potential: the
electron drop model, and the model which we call combined. The former
is simple and uniform while the latter is more accurate. Their
parameters are listed in Table \ref{tab}.

\begin{center}
\begin{table}
\caption{Parameters of electron drop and combined models} \label{tab}
\renewcommand{\arraystretch}{1.3}
\begin{tabular}{c c c c }
\hline  Model & $b_0$ & $b_2$ & $b_4$  \\
\hline \hline
Electron drop &  Eq.~(\ref{e:b0}) & Eq.~(\ref{e:b2})  & Eq.~(\ref{e:b4})  \\
Combined      &  Eq.~(\ref{e:b0_MC}) & Eq.~(\ref{e:b2})  & Eq.~(\ref{e:b4})  \\
\hline 
\end{tabular}
\end{table}
\end{center}

In the electron drop model, $b_0$, $b_2$ and $b_4$  depend on
$z=Z_2/Z_1$ but are independent of density and temperature. These
coefficients are given by Eqs.\ (\ref{e:b0})--(\ref{e:b4}). The main
disadvantage of this model is that $b_0$ is actually a slowly varying
function of the Coulomb coupling parameter which is neglected.

In the combined model, $b_0$ is given by Eq.\ (\ref{e:b0_MC}), while
$b_2$ and $b_4$ are again given by Eqs.\ (\ref{e:b2}) and
(\ref{e:b4}). Equation (\ref{e:b0_MC}) is based on extensive MC
simulations and is, therefore, more accurate than (\ref{e:b0}); Eq.\
(\ref{e:b2}) provides a robust value of $b_2$ (Sec.\ \ref{s:poten});
while $b_4$ is relatively unimportant, so that the use the electron
drop value (\ref{e:b4}) is sufficiently accurate.

\begin{figure}[t]
\centering
\includegraphics[width=0.45\textwidth]{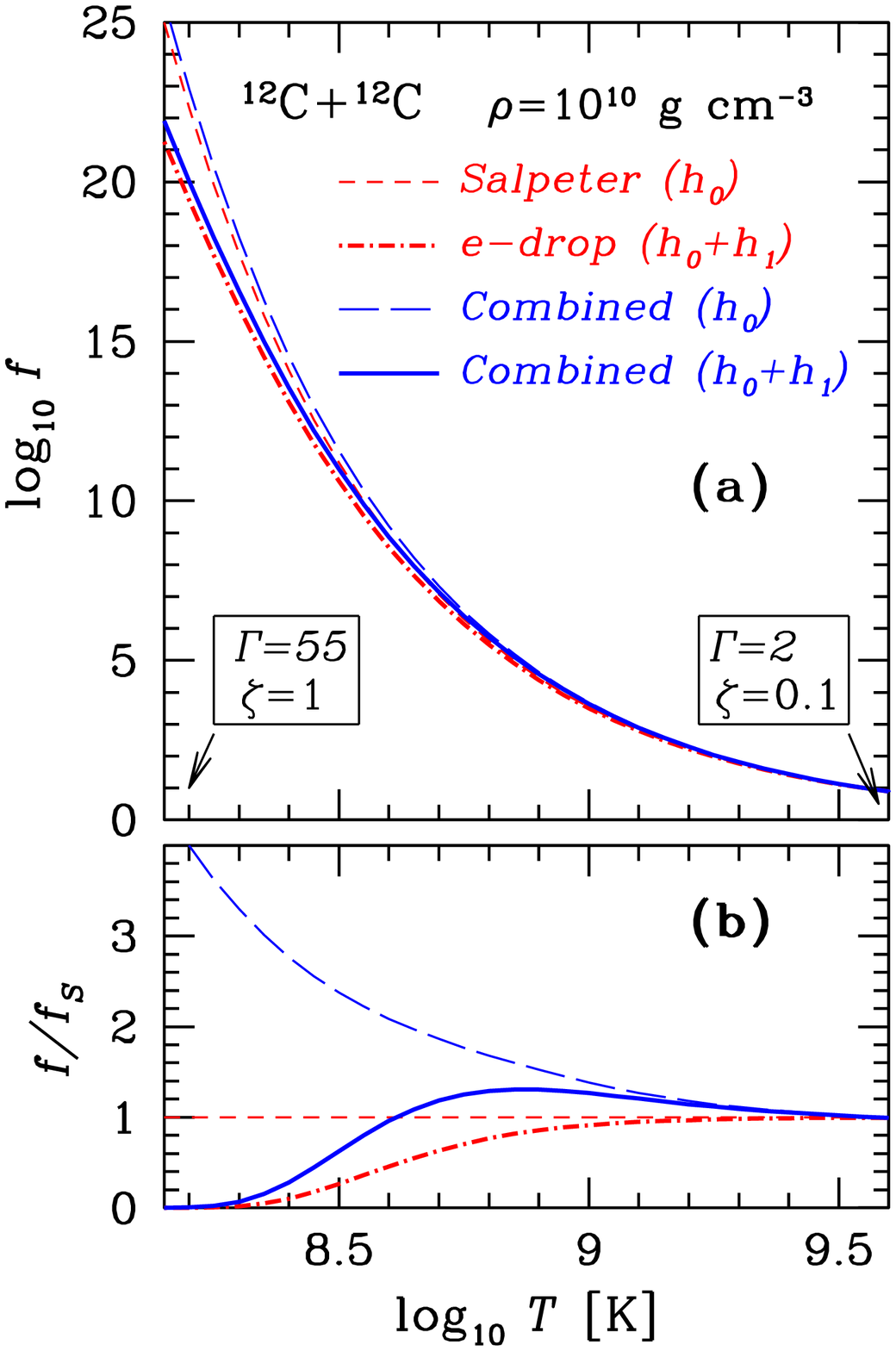}
\caption{(Color online) Enhancement of the $^{12}$C+$^{12}$C
thermonuclear reaction rate in carbon matter at $\rho=10^{10}$
g~cm$^{-3}$ as a function of temperature. {(a):} Logarithm of the
enhancement factor in the Salpeter model [short-dashed line, only
$h_0$ included in Eq.\ (\ref{e:f_scr})],  the electron drop model
(dot-dash line, $h_0$ and $h_1$ included), and in the combined model
including and excluding $h_1$ (solid and long-dashed lines,
respectively). The numbers give the values of $\Gamma=\Gamma_{12}$
and $\zeta$ at the lowest and highest temperatures displayed. {(b):}
The same enhancement factors divided by the Salpeter's factor in
natural scale (see text for details).} \label{fig:finalenh}
\end{figure}

In Fig.\ \ref{fig:finalenh} we illustrate the efficiency of the
plasma screening as a function of temperature for the
$^{12}$C+$^{12}$C reaction in carbon matter at $\rho=10^{10}$
g~cm$^{-3}$ (also see Fig.\ \ref{fig:thermorate} to understand how
this efficiency affects the reaction rate). At the highest displayed
temperature, we have $\log_{10} T$[K]=9.6, $\Gamma_{12}\approx2$ and
$\zeta\approx 0.1$, while at the lowest temperature $\log_{10}
T$[K]=8.15, $\Gamma_{12}\approx 55$ and $\zeta\approx 1$.

We present the four curves which correspond to different models and
approximations. The thin short-dashed curve refers to the Salpeter's
model where only the $h_0$ term is included in Eq.\
(\ref{e:f_scr}), with $b_0$ given by Eq.~(\ref{e:b0}). This is
equivalent to using Eq.\ (\ref{e:ffin}), where the $b_2$ and $b_4$
terms are dropped. The thick dot-dashed curve is for the full
electron drop model [$h_0$ and $h_1$ terms are included in Eq.\
(\ref{e:f_scr}); all terms are included in Eq.\ (\ref{e:ffin})]. The
thin long-dash curve is similar to the Salpeter's model but $b_0$ is
given by Eq.~(\ref{e:b0_MC}). The thick solid curve is the full
combined model [all terms included in Eq.\ (\ref{e:f_scr}) or
(\ref{e:ffin})].

Figure \ref{fig:finalenh}(a) shows logarithm of $f$ while Fig.\
\ref{fig:finalenh}(b) presents $f/f_{S}$ in natural scale for the
four approximations. When the temperature drops to $\log_{10}T=8.15$
the plasma screening enhancement intensifies and exceeds 20 orders of
magnitude. It greatly slows down the decrease of thermonuclear
reaction rate (cf.\ Fig.\ \ref{fig:thermorate}). As long as $\zeta
\ll 1$ ($T \gtrsim 10^9$~K, for a given $\rho=10^{10}$ g~cm$^{-3}$),
the screening enhancement is mainly provided by the Salpeter's term
($h_0$). However, at lower $T$ the correction $h_1$ becomes
progressively more important. It suppresses the Salpeter's screening
enhancement (by about 3 orders of magnitude at $\log_{10}T\approx
8.15$ in Fig.\ \ref{fig:finalenh}). In this way it does not allow the
plasma screening enhancement at low temperatures to become too strong
(otherwise the reaction rate would start growing up with the
temperature fall). It looks as if the $h_1$ correction
``anticipates'' the onset of the pycnonuclear burning regime at lower
temperatures and ``prepares'' the temperature-independence of the
reaction rate in the pycnonuclear regime.
This independence of temperature at lowest temperatures of
thermonuclear burning has been obtained earlier by Chugunov and
DeWitt \cite{Chugunov09} (who calculated the plasma screening
enhancement factors based on extensive MC calculations of screening
potentials).

Neglecting the $h_1$ term, we would obtain [Fig.\
\ref{fig:finalenh}(b)] a factor of four difference of the enhancement
factors in the Salpeter's and combined models at the lowest $T$.
Although this difference can be regarded as substantial, it is
actually not very important because of two reasons. First, the
enhancement factors become huge by themselves [and the difference by
a factor of four is really insignificant; see Fig.\
\ref{fig:finalenh}(a)]. Second, the inclusion of the $h_1$ correction
strongly affects the plasma screening enhancement at lowest
temperatures and reduces this difference. Therefore, even if the
combined screening model is more accurate than the electron drop one,
the difference does not seem important for applications.

Our final result, Eq.\ (\ref{e:ffin}), is obtained neglecting the
energy dependence of the astrophysical factor $S_0$. This
approximation corresponds to $f_{Sfact}=1$ in Eq.\
(\ref{e:f-decomposition}). We have checked the validity of this
assumption for the $^{12}$C+$^{12}$C reaction at $\rho=10^{10}$
g~cm$^{-3}$ as an example, by using Eq.\ (\ref{e:fSfact}) for
calculating $f_{Sfact}$ and Eq.\ (\ref{e:deltaEpk}) for finding the
shift of the Gamow peak energy. This shift appears relatively small;
$f_{Sfact}$ stays smaller but fairly close to 1 for all temperatures
displayed. The maximum difference (of about 5 per cent) from
$f_{Sfact}=1$ occurs at the lowest temperature $\log_{10}T=8.15$
shown in Fig.\ \ref{fig:finalenh}. Therefore, $f_{Sfact}=1$ is,
indeed, a good approximation.

Let us stress that nuclear reaction rates in stellar matter are also
uncertain due to rather poor knowledge of the factors $S_0$ at low
energies of astrophysical interest; see, e.g., Ref.\
\cite{Jacovlev12} and references therein. These factors are difficult
to calculate accurately and to measure in laboratory; they may have
resonant behavior which requires special consideration (even if the
strengths and positions of resonances were known -- which is usually
not the case -- see, e.g., Refs.\
\cite{aguilearaetal06,spillaneetal07} which analyze the
$^{12}$C+$^{12}$C reaction at low energies).

We add that we have studied plasma screening effects assuming rigid
(incompressible) electron background. The corrections due to finite
polarizability of the electron gas slightly increase the screening
enhancement (see, e.g., Refs.\ \cite{yasha,pc12} and references
therein) which we neglect here.

All these factors prevent exact calculation of thermonuclear reaction
rates in dense stellar matter. However, in many cases the exact rates
are not vitally important because of the strong temperature
dependence of the rates -- their uncertainties are easily
absorbed by small temperature variations making modeling of
nuclear burning phenomena almost unchanged (see, e.g., Ref.\
\cite{gasquestetal07}).

We have not focused on plasma screening of thermonuclear reactions in
the regime of weak Coulomb coupling ($T \gtrsim T_l$ in Fig.\
\ref{fig:diag}). One can use our formulae in this regime;
they would predict weak plasma screening, $f \to 1$. Although such a
description of weak screening is approximate, it should not noticeably
affect physical results. More reliable plasma screening corrections
for weak and moderate Coulomb coupling can be obtained using recent
advances in this field \cite{potekhinetal09,potekhinetal09a,chugunov12}.

At low temperatures ($\zeta \gtrsim 1$, $T\lesssim 0.34\,T_p$ in
Fig.\ \ref{fig:diag}) the present results [Eq.\ (\ref{e:ffin})]
become invalid because of the onset of the pycnonuclear burning
regime. Pycnonuclear burning has been studied in many publications
(e.g., Refs.\ \cite{svh69,Chugunov07,Chugunov09} and references
therein) but we believe that the theory of pycnonuclear reactions
is still far from being complete.

\section{Conclusions}
\label{s:conclusions}

We have developed a simple model for plasma screening in
thermonuclear reactions in dense stellar matter. In this model, the
screening is produced by an electron cloud around the reacting nuclei.
The cloud's charge is
assumed to compensate the charge of the reactants and the cloud's
shape corresponds to the minimum electrostatic energy of the reacting
system (two nuclei + electron cloud). This model is the extension of
the well known Salpeter's model of ion spheres \cite{Salp}.

The electron drop model is based on the mean field screened Coulomb
potential $H(r)$ which has been calculated for different nuclear charge
numbers $Z_1$ and $Z_2$ (Sec.\ \ref{s:poten}). We have analyzed the
low-$r$ expansion of $H(r)$ in powers of $r^2$ and derived (for the
first time) the $r^4$ expansion term, Eq.\ (\ref{e:b4}). We have
computed $H(r)$ and obtained a simple fit, Eq.\ (\ref{e:hfit}). In
passing, we have calculated $H(r)$ for the model in which the electron
cloud is approximated by a prolate Maclaurin ellipsoid (Appendix
\ref{s:maclaurin}).

At the next step we have added the plasma screening Coulomb potential
to the total effective potential $U_\mathrm{eff}(r)$, which governs
nuclear reaction. With this new potential we have calculated the
astrophysical $S$ factors (Sec.\ \ref{s:sfact}). Our generalized
$S$ factors include the effects of nuclear and plasma screening
interactions on the same footing. In the regime of strong Coulomb
coupling of atomic nuclei in dense matter, they depend on the density
of the matter, $\rho$. We have analyzed the properties of such $S$
factors and presented simple analytic approximations.
These generalized $S$ factors are strongly modified by plasma
screening at low energies $E$ and high densities $\rho$. For
``ordinary'' nuclear reactions like $^{12}$C+$^{12}$C the inclusion
of plasma screening into $S$ factors is equivalent to introducing
traditional plasma screening enhancement factors in the reaction
rates. However generalized $S$ factors would be more natural for
those reactions for which effective plasma screening length is
comparable to sizes of the reacting nuclei. Such reactions could
occur at high temperatures in the inner crust of neutron stars;
nuclear and plasma physics effects would be not separated there.

Finally, we have used the electron drop model and calculated
thermonuclear reaction rates, studied their plasma screening
enhancement and approximated the enhancement factors by analytic
expressions (Sec.\ \ref{s:rate}). We have analyzed the properties of
plasma screening enhancement and proposed a combined analytic model
for strong plasma screening in thermonuclear reactions (Sec.\
\ref{s:discussion}).

Our results  are in good agreement with those obtained by other
techniques. The advantage of our model is that it is well formulated,
physically transparent, and easily formalized in terms of analytic
approximations at every step of investigation (for the effective
potentials, astrophysical $S$ factors, and enhancement factors of
nuclear reactions). The results can be used to model various
astrophysical manifestations of nuclear burning and nucleosynthesis
in white dwarfs and neutron stars.

Strictly speaking, our results cannot be used in the thermonuclear
regime with weak plasma screening (high temperatures, low densities)
but in that case the plasma screening has almost no effect on the
reaction rates. Equally, the results are inapplicable at very low
temperatures and high densities, where the pycnonuclear effects
(zero-point vibrations of the reacting nuclei) become pronounced.

So far almost all the calculations of the reaction rates in dense
matter have been performed within the mean field potential. However,
the plasma potential created by neighboring plasma particles
is actually fluctuating (depends of specific configuration of neighboring
particles). Our model can be generalized to the case of fluctuating
potential by introducing an ensemble of electron drops (of different
shapes) around the reactants and probabilities of their realizations.
This has perspective to study the effect of plasma field fluctuations
on the reaction rates in dense matter.

\begin{acknowledgements}
We are grateful to H.\ DeWitt and A.\ Chugunov for useful comments. DGY acknowledges
partial support from RFBR (Grants No. 14-02-00868 and No. 13-02-12017-ofi-M) and RF Presidental Program NSh 294.2014.2.
\end{acknowledgements}

\appendix

\section{Coulomb energy}
\label{sec:pert}

In the Appendices we show how to derive the first three coefficients
$b_0$, $b_2$ and $b_4$ in the small-$x$ expansion~\eqref{e:lowxexp}
of the screening potential in the electron drop model (Sec.\
\ref{s:model}). To obtain these coefficients the shape of the
electron drop can be approximated by a prolate Maclaurin ellipsoid
for which the electrostatic problem is solved analytically (see,
e.g., Ref.~\cite{Chandra}). First, we define the main quantities,
then present the equations for the Maclaurin ellipsoid
(Appendix~\ref{s:maclaurin}) and build the desired expansion
(Appendix~\ref{s:hfinal}). Here we use standard physical units.

At $r\leq (a_1+a_2)$ the Coulomb potential $U_C(r)$ in Eq.\
(\ref{e:screenpot}) can be written as
\begin{equation}
  U_C(r)=\frac{Z_1Z_2e^2}{r}+W(r)-W_{12},\quad
  W(r)=W_{ee}(r)+W_{ei}(r),
\label{e:AU_C}
\end{equation}
where $W(r)$ is the electrostatic energy of the drop (excluding
direct Coulomb interaction of point-like ions). It contains the
interaction energy of the ions (positioned at $\bm{r}_1$ and
$\bm{r}_2$) with the electron drop
\begin{equation}
   W_{ei}(r)=Z_1e\Phi(\bm{r}_1)+Z_2e\Phi(\bm{r}_2),
\label{e:AU_ei}
\end{equation}
and the electrostatic energy of the drop
\begin{equation}
    W_{ee}(r)=-\frac{e n_e}{2} \int_V dV\, \Phi(\bm{r}).
\label{e:AU_ee}
\end{equation}
In this case
\begin{equation}
    \Phi(\bm{r})=-en_e \int_V \frac{dV'}{|\bm{r}-\bm{r}'|}
\label{e:APhi}
\end{equation}
is the electrostatic potential created by the drop. The integration
in Eqs.~(\ref{e:AU_ee}) and (\ref{e:APhi}) is carried over the volume
$V$ of the electron drop confined within the surface $\partial
V=S(r)$. The term $W_{12}$ in Eq.\ (\ref{e:AU_C}) is given by Eq.\
(\ref{e:U12}); it is introduced to satisfy the condition $U_C(r)=0$
at $r\geq (a_1+a_2)$ (Sec.\ \ref{s:model}).

At $r\leq (a_1+a_2)$ the screening energy $H(r)$  in
Eq.~(\ref{e:screenpot}) is given by
\begin{equation}
  -H(r)=W(r)-W_{12}. \label{e:uc_def}
\end{equation}

\section{Maclaurin ellipsoid model}
\label{s:maclaurin}

At small $r$ we approximate the electron drop by a Maclaurin
ellipsoid prolate along the $z$-axis. The radial coordinate
$r_s(\theta)$ of the surface $S$ is then given by
\begin{equation}
   \frac{1}{r_s^2(\theta)}=\frac{\cos^2 \theta }{a_\parallel^2}
   +\frac{\sin^2 \theta }{a_\perp^2},
\label{e:Ashape}
\end{equation}
where $a_\parallel$ and $a_\perp$ are the ellipsoid semi-axes (along
the $z$-axis and in the perpendicular plane, respectively), and
$\theta$ is the polar angle.

The electron charge within the ellipsoid should compensate the charge
of the ions. This gives
\begin{equation}
    Z_c=\frac{4\pi}{3}a_\parallel a_\perp^2n_e=\frac{a_\parallel
    a_\perp^2}{a_e^3};
\label{e:AVol}
\end{equation}
$a_e$ and $Z_c$ are defined in Sec.\ \ref{s:param}. Introducing the
ellipticity of the ellipsoid,
$\epsilon=\sqrt{1-(a_\perp/a_\parallel)^2}$, we have
\begin{equation}
  a_\parallel=a_c\,(1-\epsilon^2)^{-1/3},
  \quad a_\perp=a_c\,(1-\epsilon^2)^{1/6},
\label{e:Aa1a2}
\end{equation}
where $a_c$ is given by Eq.~(\ref{a12}).

The potential $\Phi(\bm{r})$ within the ellipsoid is
\begin{equation}
   \Phi(\bm{r})=-\pi e n_e (I-A_\parallel z^2-A_\perp x^2 - A_\perp
   y^2),
\label{e:AAPhi}
\end{equation}
with
\begin{equation}
  A_\parallel=2 \frac{1-\epsilon^2}{\epsilon^2}\,(L-1),\quad
  L=\frac{1}{2\epsilon}\,\ln\left( 1+\epsilon \over 1-\epsilon \right),
\label{e:AAL}
\end{equation}
$I=2 a_\perp^2 L$, and $A_\perp=1-A_\parallel/2$. Integrating the
potential $\Phi(\bm{r})$ one obtains the Coulomb energy of the
electron drop, 
\begin{equation}
   W_{ee}=\frac{3}{5} \left(  4 \pi \over 3 \right)^2
   e^2n_e^2 a_\perp^4 a_\parallel L.
\label{eAAUee}
\end{equation}

In our case the ions are placed at the $z$-axis, so that
\begin{equation}
  W(r)=-\pi e^2 n_e \left[Z_cI-Z_1A_\parallel z_1^2-Z_2A_\parallel
  z_2^2\right]+W_{ee}.
\label{e:AUUc}
\end{equation}
These formulae give $U_C(r)$ for any electron drop in the form of
Maclaurin ellipsoid.

\section{Small-$r$ expansion}
\label{s:hfinal}

At $r \ll (a_1+a_2)$ the electron drop shape is well
approximated by an ellipsoid with ellipticity $\epsilon \ll 1$. From
Eq.~(\ref{e:Ashape}) the shape becomes
\begin{equation}
  r_s(\theta)=a_c \,
  \left(1+\frac{1}{3}\,{\epsilon^2 P_2(\cos \theta)}\right),
  \label{e:shape}
\end{equation}
where $P_2(x)={1\over 2}(3x^2-1)$. Higher-order terms are unimportant
for our problem.

In the expression for the electrostatic energy, we expand
$A_\parallel$, $I$, and $W_{ee}$ in powers of $\epsilon^2$ keeping
first- and second-order terms,
\begin{align}
  A_\parallel&={2 \over 3} - {4\over 15}\,\epsilon^2-{4 \over
  35}\,\epsilon^4, \\
  I&=2 a_c^2 \left(1-\frac{1}{45}\, \epsilon^4\right), \\
  W_{ee}&=\frac{4\pi}{5}Z_ce^2n_e\,a_c^2
  \left(1-\frac{1}{45}\, \epsilon^4\right).
\end{align}

Placing the coordinate origin in the center of the electron drop, we
have
\begin{equation}
  z_1=-{Z_2r}/{Z_c},\quad
  z_2={Z_1r}/{Z_c}. \label{e:z1z2}
\end{equation}
Now $\epsilon$ is the only free parameter. We will
see that $\epsilon\propto r$; accordingly we expand $W(r)$ keeping
the terms $r^4,\epsilon^4,\epsilon^2 r^2$:
\begin{align}
    W(r)=&2\pi a_c^2e^2n_e
    \left[-\frac{3Z_c}{5}+\frac{3Z_c\epsilon^4}{225}\nonumber\right.\\
    &\left.+\frac{Z_1Z_2x^2}{3Z_c}
    -\frac{2Z_1Z_2x^2\epsilon^2}{15 Z_c}\right],
\end{align}
where $x=r/a_c$. The optimal value of $\epsilon$ is found by
minimizing $W(r)$,
\begin{equation}
    \epsilon^2=5Z_1Z_2x^2/Z_c^2.
    \label{e:ellipticity}
\end{equation}
At this $\epsilon$,
\begin{equation}
  W(r)=\frac{e^2}{a_e} \left[-0.9\,Z_c^{5/3}+
  \frac{Z_1Z_2}{2Z_c}\left(\frac{r}{a_e}\right)^2
  -\frac{Z_1^2Z_2^2}{2Z_c^{11/3}}\left(\frac{r}{a_e}\right)^4\right].
  \label{e:uae}
\end{equation}
Substituting this into Eq.~(\ref{e:AU_C}), we reproduce
Eq.~\eqref{e:lowxexp}.

We have derived the interaction energy for a specific choice of the
ellipsoid center; it satisfies the minimum energy requirement with
respect to variations of $\epsilon$. Now we sketch the proof that
this solution is exact up to the $r^5$ order, and minimizes energy
with respect to any perturbations of the drop's shape.

Note that the minimum energy requirement is equivalent to the
condition of constant (zero) total potential on the $S$ surface.
We have explicitly checked that this condition is satisfied.
Moreover, we find that the potential induced by the ion charges and
the {\em unperturbed} electron spherical drop on the surface of the
{\em exact perturbed} electron drop is $\sim r^2$. It is important
that the shape is spherical in the $r$ order (with our
specific choice of the coordinate origin). Therefore the deviation of
the perturbed surface from the unperturbed one is $\sim r^2$. Since
the potential mentioned above is $\sim r^2$, the shape correction
$\sim r^3$ contributes only to the energy corrections $\sim r^5$ and
higher. However, it is easily seen that there are no odd-order energy
terms; hence the expansion is correct up to the terms $\sim r^5$. The
absence of the odd-order terms is intuitively obvious, because all
the intermediate equations respect the symmetry transformation
$r\rightarrow -r,\;\theta\rightarrow\pi-\theta$, and $\theta$ is then
integrated out in order to obtain the energy. Note that this symmetry
is consistent with the Widom expansion~\cite{widom63}.



\end{document}